\documentclass[10pt]{iopart}
\usepackage{iopams}
\usepackage{cite}
\usepackage{graphicx}
\usepackage{xcolor}

\begin{document}
\title{Perfect transmission of 3D massive Kane fermions in HgCdTe Veselago lenses}

\author{Y Betancur-Ocampo and V Gupta}
\address{Departamento de F\'isica Aplicada, Centro de Investigaci\'on y de Estudios Avanzados del IPN, A.P. 73 Cordemex 97310 M\'erida, Yucat\'an, M\'exico}
\ead{yonatan.betancur@gmail.com}

\begin{abstract}
The transmission properties of three-dimensional (3D) massive Kane fermions in HgCdTe (MCT) heterojunctions have been studied using the simplified Kane-Mel\'e model. Based on our theoretical calculations, we propose the design of an electronic device, called mass inverter, which consists of the junction of a narrow-gap semiconductor and semimetal. Such a device can be used in electron optics applications, since it operates as a Veselago lens and presents Klein tunneling (KT) of 3D massive Kane fermions under normal incidence. We found that KT and Veselago lensing can also be observed for general MCT heterojunctions with a specific value of doping level. We show that non-resontant perfect transmission of massive Kane fermions persists in a potential barrier for heterojunctions formed by a semimetal between two standard semiconductors. This effect is quite robust when the ideal conditions of a possible experimental test are deviated. Our findings may have important implications in the development of nano-electronic devices using 3D massive Kane fermions, where transmission features of massless particles are recovered.
\end{abstract}

\submitto{\JPCM}
\vspace{2pc}

\maketitle


\section{Introduction} \label{intro}
 
Relativistic material is a recent condensed matter concept for classifying a wide variety of systems whose excitations have a dispersion relation forming Dirac cones \cite{Wehling}. There are famous physical realizations, such as graphene and several topological insulators, where electrons behave like massless Dirac fermions in a two-dimensional (2D) space \cite{Novoselov,Castro,XLQi}. The outstanding properties of these materials, as negative refraction of massless Dirac fermions, have been recently observed \cite{Ho Lee, Chen,Cheianov}. The possibility of relativistic materials presenting excitations with higher dimensionality and pseudo-spin value has been explored in order to find novel and singular physics \cite{Borisenko,Liu,Neupane,Orlita,Teppe,Nicol,Lan,Bercioux2,Essafi,Betancur,Bercioux,Louie,Shen,Carusotto,Kusmartsev, Grushin}. This exploration has begun to give amazing discoveries in pseudo-spin one systems, such as the realization of localized states \cite{Vicenzio,Muk,Apaja}, the prediction of room-temperature superconductivity \cite{Peotta,Peotta2}, and the super-Klein tunneling of massless and massive particles \cite{Louie,Bercioux,Shen,Betancur}. These phenomena are due mainly to the appearence of a flat band in the energy band structure. 

Klein tunneling (KT) is perhaps the most outstanding electron transmission phenomena discovery in graphene, where massless Dirac fermions cross an arbitrary potential barrier without backscattering \cite{Klein,Katsnelson,Kim,Peres,Libisch,Wilmart}. Currently, this effect continues emerging in other relativistic materials with enlarged pseudo spin structure \cite{Bercioux,Shen,Louie,Ghosh2}. Moreover, massive pseudo-spin one particles present omnidirectional perfect transmission in Lieb lattices and other 2D systems \cite{Betancur}. Such an effect has been claimed to have possible applications in electron quantum optics \cite{Bercioux,Shen,Louie,Betancur}.

Recently, Hg$_{1-\chi}$Cd$_{\chi}$Te (MCT) crystals have been identified as a three-dimensional (3D) relativistic materials. Their electronic band structure also present Dirac cones, where a flat band touches one dispersive band \cite{Harman,Groves,Guldner,Bernevig,Konig,Orlita,Teppe}. Electrons in MCT crystals, called Kane fermions \cite{Orlita}, have a pseudo-relativistic behavior with atypical pseudo-spin structure. The conservation of this pseudo-spin allows the emergence of KT for massless Kane fermions, whose dynamics can be depicted by a simplified Kane Hamiltonian \cite{Orlita,Teppe,Nicol,Du}. These kind of crystals promise to have unusual physics which is being actively addressed. 

Although important experimental contributions in MCT crystals have been realized \cite{Orlita,Teppe}, the transmission properties of 3D massive Kane fermions continue unexplored yet. The fact that 3D massive Kane fermions have an angular momentum structure different from 2D massive pseudo-spin one particles, it opens up the possibility of unusual transport phenomena. MCT heterostructures may have concrete applications in nano-electronics by its tunable band gap energy. Thus, in the present work we theoretically study the transmission properties of 3D massive Kane fermions in $n$-$p$ and $n$-$p$-$n$ heterojunctions of MCT crystals. We found that perfect transmission appears for normal incidence when the sign of the effective mass is inverted in the interface. Non-resonant perfect transmission and focusing of electron flow are obtained for specific values of doping level in a general heterojunction. Such systems offer the opportunity of restoring transmission features of massless fermions. Further, the robustness of the perfect transmission of 3D massive Kane fermions suggests that the experimental observation might be feasible.

The present work is organized as follows. In the second section we apply the simplified Kane-Mel\'e model for describing the dynamics of 3D massive Kane fermions in MCT crystals. In the third section we study  the scattering of particles and the conservation of pseudo-spin for an $n$-$p$ heterojunction. We analyze the transmission features of 3D massive Kane fermions in the mass inverter, which is a device changing the sign of the mass in the interface, as shown in fourth section. The effect of a second interface forming a $n$-$p$-$n$ heterojunction is exhibited in the fifth section, where we observe how the Fabry-P\'erot resonances modifies the transmission probability. Moreover, we show that the perfect transmission of 3D massive Kane fermions is quite robust when the ideal conditions are deviated. Our conclusions and final remarks are presented in last section.
 
\section{Simplified Kane-Mel\'e model} 

For describing the band structure of MCT crystals, a standard Kane-Mel\'e model is employeed in the study of massive Kane fermions dynamics \cite{Orlita,Nicol,Teppe,Kane}. It considers an 8 $\times$ 8 Kane Hamiltonian, which is able to obtain the appropriate energy bands evolution by varying the Cd concentration and temperature \cite{Harman,Groves,Guldner}. The band energy region of interest is located near the centre of the first Brillouin zone, where two dispersive ($\Gamma_6$ and $\Gamma_8$) bands and one heavy hole band present two-fold degeneracy for low energy regime (around hundreds of meV). The critical value of $\chi_c \approx 0.17$ at $T \approx 0$ K, corresponding to massless Kane fermions, separates the semiconductor phase ($\chi > \chi_c$) of semimetal ($\chi < \chi_c$). In this topological phase transition, the $\Gamma_6$ and $\Gamma_8$ bands are interchanged \cite{Orlita}. The particular shape of these energy bands resembles the dispersion relation of massive pseudo-relativistic particles with spin one. Thus, the essential physics of charge carriers in MCT crystals can be obtained from a simplified Kane model \cite{Orlita, Nicol, Teppe}. It is obtained when the quadratic terms of the linear momentum are neglected and the influence of the split-off $\Gamma_7$ band is much reduced due to the large splitting of $\Delta \approx 1$ eV \cite{Orlita}. Under these approximations, the simplified Kane model consists of a 6 $\times$ 6 Hamiltonian \cite{Teppe}

\begin{equation}
H = v_F\vec{A}\cdot\vec{p} + mv^2_F\Lambda,
\label{KH}
\end{equation}

\noindent which contains only two parameters: the Fermi velocity $v_F$, whose experimental value was estimated to be $v_F \approx 1.06 \times 10^6$ ms$^{-1}$ being a universal constant in MCT crystals \cite{Teppe}, and the effective mass $m$, which can be tunable by changing the Cd concentration $\chi$ and temperature $T$ \cite{Teppe}. The sign of the mass for the semimetal (semiconductor) phase is negative (positive). The linear momentum in Hamiltonian \eref{KH} is $\vec{p} = -i\hbar \vec{\nabla}$. While the 6 $\times$ 6 matrices $\vec{A} = (A_x,A_y,A_z)$ and $\Lambda$ 

\begin{eqnarray}
   A_x &= \sigma_z\otimes J_x, &
   A_y= \mathbb{I}\otimes J_y,\nonumber \\
   A_z &= \sigma_x\otimes J_z, \quad
 \textrm{and} \quad &
   \Lambda = \mathbb{I}\otimes K
\end{eqnarray}

\noindent are given as the tensorial product of the $2\times2$ identity matrix $\mathbb{I}$, the Pauli matrices $\sigma_i$ where $i = x \: \textrm{or} \:z$, the mass generator term in MCT crystals 

\begin{equation}
   K = \left(
  \begin{array}{ccc}
    1 & 0 & 0\\
    0 &-1 & 0\\
    0 & 0 &-1
  \end{array}\right),
\end{equation} 

\noindent and the $3\times3$ angular operator $\vec{J} = (J_x,J_y,J_z)$ defined by the matrices 

\begin{eqnarray*}
   J_x = \frac{1}{2}\left(
  \begin{array}{ccc}
    0 & \sqrt{3} & -1\\
 \sqrt{3} & 0 & 0\\
    -1 & 0 & 0
  \end{array}\right), \\ 
   J_y= \frac{i}{2}\left(
  \begin{array}{ccc}
    0& \sqrt{3} &1 \\
    -\sqrt{3} & 0 & 0\\
    -1 & 0 & 0
  \end{array}\right),
\end{eqnarray*}

\begin{equation}
\textrm{and}\quad
J_z= \left(\begin{array}{ccc}
    0&0&-1 \\
    0&0&0\\
   -1&0&0
  \end{array}\right).
\end{equation}

\noindent The components of $\vec{J}$ do not satisfy the algebra of angular momentum for spin-one. Hence, the behavior of electrons in MCT crystals is different to massive pseudo-spin one particles. For that reason, these excitations are called as Kane fermions \cite{Orlita}. Nevertheless, Kane fermions share common physical properties with relativistic particles \cite{Orlita}. 

\noindent When we consider $p_z = 0$, the Kane Hamiltonian in equation \eref{KH} can be represented by a 6 $\times$ 6 block diagonal matrix where one 3 $\times$ 3 block is written as

\begin{equation}
H_s = v_F \vec{J}\cdot \vec{p} + mv^2_F K,
\label{SKH}
\end{equation}
 
\noindent while the other one is $H'_s = -v_F J_xp_x + v_F J_yp_y + mv^2_F K$. Using the expression \eref{SKH} for solving the secular problem, the relativistic dispersion relation is given by

\begin{equation}
E_0 = -mv^2_F,
\: \: \: \: \: \: \: \: \:
E_s = s\sqrt{m^2v^4_F + v^2_Fp^2},
\label{dr}
\end{equation}

\noindent where $E_0$ corresponds to a dispersionless heavy hole band, the band index $s = 1$ $(-1)$ indicates the conduction (valence) band. We note that the flat band $E_0 = -mv^2_F$ is located at the top (bottom) of valence (conduction) band for the semiconductor (semimetal) phase, as shown in \fref{fig2}(b). The energy bands in equation \eref{dr} are again obtained using the block $H'_s$. Thus, the bands are twofold degenerate due to the time-reversal symmetry. Then, by Kramer theorem, there is a valley pseudo-spin 1/2 structure whose doublet can be represented by $|\uparrow\rangle$ and $|\downarrow\rangle$. Valley-independent physical properties are identically described either by $H_s$ or $H'_s$. These valleys, which can be denoted as $\Gamma$ and $\Gamma'$, are located at the centre of the first Brillouin zone. The corresponding eigenstates of $H_s$ have the form

\begin{equation}
|\Psi_0\rangle = \frac{1}{2}\left(\begin{array}{c}
0\\
\textrm{e}^{-i\phi}\\
\sqrt{3}\textrm{e}^{i\phi}
\end{array}\right),
\: \: \: \: \: \: \: \: \:
|\Psi_s\rangle = \frac{1}{2\sqrt{2}}\left(\begin{array}{c}
2s\beta\\
\sqrt{3}\gamma \textrm{e}^{-i\phi}\\
-\gamma \textrm{e}^{i\phi}
\end{array}\right)
\label{WM1}
\end{equation}

\noindent where $\gamma = 1 - (smv^2_F/E)$, $\beta = [1 - (mv^2_F/E)^2]^{1/2}$, and $E = [(mv^2_F)^2 + v^2_Fp^2]^{1/2}$ is the particle energy. The wavefunction's phase $\phi$ is related to the propagation direction since $\tan(\phi)= p_y/p_x$ and it is geometrically interpreted as the angle formed by the particle beam and the $x$-axis. For $p_z = 0$, the complete base of six elements of the Kane Hamiltonian in equation \eref{KH} is obtained by the tensorial product of $|\uparrow\rangle$ ($|\downarrow\rangle$) with $|\Psi_0(\phi)\rangle$ ($|\Psi_0(\pi+\phi)\rangle$) and $|\Psi_s(\phi)\rangle$ ($|\Psi_s(\pi+\phi)\rangle$).

 The specific location of the flat band and pseudo-spin structure cause drastic effects on the transmission probability \cite{Betancur}. Therefore, the transmission of 3D massive Kane fermions must have unique features which are distinctives regarding 2D massive pseudo-spin one particles.

\section{Transmission through MCT hererojunctions}

\begin{figure}
\begin{tabular}{l}
(a)\qquad \qquad \qquad \qquad \qquad \qquad \qquad \qquad  \qquad \qquad \qquad \qquad\\
\includegraphics[trim = 0mm 0mm 0mm 0mm, scale = 0.41]{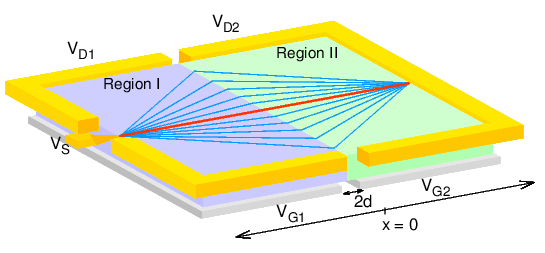}\\
(b)\qquad \qquad \qquad \qquad \qquad \qquad \qquad \qquad  \qquad \qquad \qquad \qquad\\
\includegraphics[trim = 0mm 0mm 0mm 0mm, scale = 0.46]{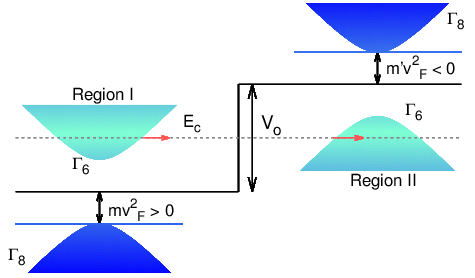}
\end{tabular} 
\caption{(a) Description of the particle scattering through the MCT heterojunction; the particle flow (blue lines) is injected from the source $V_S$ at the left of the region I. The mass inverter, which is tuned at the energy $E = V_o/2$, operates as a Veselago lens by focusing the refracted flow to a spot in the region II. The Red line indicates the KT of massive Kane fermions (2D case is shown). The gates $V_{G1}$ and $V_{G2}$, which are separated by a distance $2d$, create the step potential and dope the narrow-gap semiconductor and semimetal. $V_{D1}$ and $V_{D2}$ are located on both sides for draining the reflected and refracted particles. The reflected flow in the region I is not drawn. (b) Scheme of a junction depicted by a step potential $V_0$ and the corresponding energy band structure. The label $E_c$ indicates the Fermi level where the incident particle flow is focused in the region II.}
\label{fig2}
\end{figure}

We consider the heterojunction as shown in \fref{fig2}(a), which can be formed by two different MCT crystals. Two external gates are located in regions I and II creating the step potential $V_0$ and separated by a distance $2d$, where $d$ is defined as the split-gate length. The value of $d$ allows distinguish whether the step potential is abrupt under the condition $(\beta k_B d)^2 < \beta k_Fd < 1$ (see appendix A), where $k_B = mv_F/\hbar$ ($k_F = E_F\beta/\hbar v_F$) is the de Broglie (Fermi) wave vector. In a similar way to massless particles, the condition avoids the filter of particles with incidence angles beyond the normal direction \cite{Low2,Fuchs,Ghosh} as well as the decreasing of transmission probability due to the finite mass (see appendix A). Ballistic transport regime can be reached when the coherence length and mean free path of the particle are larger than the dimensions of the device. Typical experimental values of coherence length and mean free path of massive Kane fermions in MCT crystals are of the order of $\mu$m \cite{Noel,Daumer}.

In order to analyze the 3D massive Kane fermions transmission through a heterogeneous junction, a point source injects the electron flow, as shown in \fref{fig2}(a). The incident particle beam with incidence angle $\phi$ is scattered in the interface located at $x=0$ producing a reflected (refracted) beam along the $-\phi$ ($\phi'$) direction. The output particle flow is collected by an extended drain. Both reflected and refracted beams are located in the same plane defined by the $x$-axis and incidence beam. Since $p_y$ and $p_z$ are conserved due to the traslational symmetry along $y$ and $z$, it is always possible to rotate the reference system around the $x$-axis so that the new $xy$-plane matches with the incident plane. This observation allows as to set $p_z = 0$ in all the cases. In consequence, the transmission properties have azimuthal symmetry around the $x$-axis and the 3D massive Kane fermions dynamics can be described using the Hamiltonian \eref{SKH} with an abrupt step potential $V(x)$

\begin{equation}
H = v_F \vec{J}\cdot\vec{p} + mv^2_FK + V(x), 
\label{KHV}
\end{equation}

\noindent
where $V(x)$ is zero ($V_0$) for the region I (II), whose effect on the band structure is to raise the Dirac cone energy by $V_0$ in the region II, as seen in \fref{fig2}(b). Throughout the text, the unprimed (primed) quantities correspond to the region I (II). The conservation of $p_y$ and $E$ allow to establish a Snell law for relativistic particles, regarless of their pseudo-spin, namely

\begin{equation}
\phi' = (s - s')\frac{\pi}{2} + ss'\arcsin\left(\frac{|E|\beta}{|E - V_0|\beta'}\sin\phi\right).
\label{Snell}
\end{equation}

\noindent Again as in the massless pseudospin-$1/2$ case \cite{HoLee,Chen,Cheianov}, negative refraction of massive Kane fermions is obtained. This result is the same for pseudo relativistic particles due to the inversion of $\vec{p}$ for interband tunneling. Hence, Veselago lensing with pseudo relativistic particles can be implemented in heterojunctions under the general focusing condition 

\begin{equation}
E_c = V_0/2  + (m^2 - m'^2)v^4_F/2V_0,
\label{FC}
\end{equation} 

\noindent which is valid when $V_0 \geq (|m| + |m'|)v^2_F$. The substitution of equation \eref{FC} in the Snell law \eref{Snell} leads to the relation $\phi' = \pi - \phi$, indicating the presence of focused particle beams in the region II, as shown in \fref{fig2}(a). It is also interesting to note that using the focusing condition \eref{FC} and $m = - m'$, $J_x$ is conserved under normal incidence (see appendix B). This result means that massive Kane fermions present KT in MCT heterojunction formed by a semiconductor and semimetal regions. Such a heterojunction, called mass inverter, changes the sign of the mass when the Kane fermion is scattered. Thus, the particle with $E_c = V_0/2$ crosses from the conduction (valence) band in the region I to the valence (conduction) band in the region II reversing both the linear momentum and the sign of the mass for obtaining the conservation of $J_x$. This effect can also be understood noting that the mass inverter, operating as a Veselago lens \cite{Ves,Pend}, contains the same doping level and mass density with opposite sign on both sides of the junction (see \fref{fig2}(b)). When massive Kane fermion are focused the charge remains unchanged, then electric current is different from zero. However, the device inverts the mass making that the mass current becomes zero as in massless case. Thus, the conservation of $J_x$ is restored.

In order to calculate the transmission probability of Kane fermions for an MCT heterojunction, we express the wavefunction at the conduction or valence band in the region I ($x < 0$) as 

\begin{eqnarray}
|\Psi_I\rangle & = &\left(\begin{array}{c}
\psi_I^{(1)}(x)\\
\psi_I^{(2)}(x)\\
\psi_I^{(3)}(x)
\end{array}\right) \nonumber\\
& = &\frac{1}{2\sqrt{2}}\left(\begin{array}{c}
2s\beta\\
\sqrt{3}\gamma\textrm{e}^{-i\phi}\\
-\gamma\textrm{e}^{i\phi}
\end{array}\right)\exp(ixp_x/\hbar) + \nonumber \\
& &   
\frac{1}{2\sqrt{2}}r\left(\begin{array}{c}
-2s\beta\\
\sqrt{3}\gamma\textrm{e}^{i\phi}\\
-\gamma\textrm{e}^{-i\phi}
\end{array}\right)\exp(-ixp_x/\hbar),
\label{pwr1}
\end{eqnarray}

\noindent 
where $p_x=|E|\cos(\phi)/v_F$ is the component of the linear momentum in the $x$ direction. The coefficient $r$ is the probability amplitude for the reflected beam. In the region II, the transmitted wavefunction is given by

\begin{eqnarray}
|\Psi_{II}\rangle & = \left(\begin{array}{c}
\psi_{II}^{(1)}(x)\\
\psi_{II}^{(2)}(x)\\
\psi_{II}^{(3)}(x)
\end{array}\right) \nonumber \\
 & = \frac{1}{2\sqrt{2}}t\left(\begin{array}{c}
2s'\beta'\\
\sqrt{3}\gamma'\textrm{e}^{-i\phi'}\\
-\gamma'\textrm{e}^{i\phi'}
\end{array}\right)\exp(ixp'_x/\hbar), \nonumber \\
 & 
\label{pwr2}
\end{eqnarray}

\noindent where $\gamma' = 1 - s'm'v^2_F/|E - V_0|$, $\beta' = [1 - m'^2v^4_F/(E - V_0)^2]^{1/2}$, and $s'= \textrm{sgn}(E - V_0)$. The amplitude for the transmitted beam is denoted by $t$ and $p'_x=|E-V_0|\cos(\phi')/v_F$ is the linear momentum along the $x$-axis. The probability amplitudes $r$ and $t$ are determined using the boundary conditions at $x = 0$ for the wavefunctions \eref{pwr1} and \eref{pwr2}. Integrating the Kane equation with the Hamiltonian \eref{SKH} from $-\epsilon$ to $\epsilon$ and taking the limit $\epsilon \rightarrow 0$, where $m(x)$, $V(x)$ and $\psi_i(x)$ are assumed to be finite, we find that 

\begin{eqnarray}
\psi_I^{(1)}(0^-) =  \psi_{II}^{(1)}(0^+) \nonumber \\
\sqrt{3}\psi_I^{(2)}(0^-) - \psi_I^{(3)}(0^-) =  \sqrt{3}\psi_{II}^{(2)}(0^+) - \psi_{II}^{(3)}(0^+) 
\label{BC}
\end{eqnarray}

\noindent must be continuous. These boundary conditions allow to obtain the $2 \times 2$ linear equation system for $r$ and $t$

\begin{eqnarray}
s\beta(1 - r) & =  s'\beta't \nonumber \\
w^* + wr & = w'^*t,
\label{ES}
\end{eqnarray}

\noindent where $w = (3\textrm{e}^{i\phi} + \textrm{e}^{-i\phi})\gamma$ and $w'$ has identical form than $w$ in terms of $\phi'$ and $\gamma'$. The superscript * indicates complex conjugate. Solving the 2 $\times$ 2 equation system \eref{ES}, the reflection probability is

\begin{eqnarray}
R & = & |r|^2 = \nonumber \\
 &  & \frac{(s\xi\cos\phi - s'\xi'\cos\phi')^2 + (s\eta\sin\phi - s'\eta'\sin\phi')^2}{(s\xi\cos\phi + s'\xi'\cos\phi')^2 + (s\eta\sin\phi - s'\eta'\sin\phi')^2}, \nonumber \\
 & & 
\label{RC}
\end{eqnarray}

\noindent where $\xi = 2\gamma\beta'$, $\xi' = 2\gamma'\beta$, $\eta = \gamma\beta'$ and $\eta' = \gamma'\beta$. Using the equation \eref{RC}, we can confirm the perfect transmission of massive Kane fermions for $m = -m'$, $E = E_c = V_0/2$, and $\phi = 0$ showing that $R = 0$. In general, perfect transmission for normal incidence is obtained in a heterojunction when the Fermi level has the value of

\begin{equation}
E_K = \frac{m}{m-m'}V_0.
\label{EK}
\end{equation}

\noindent We note that $E_K$ in equation \eref{EK} could have values within the band gap energy. For these cases, the perfect transmission of massive Kane fermions is forbidden. If the junction has the same (opposite) sign of masses on both regions, the allowed KT occurs inside the intraband (interband) transmission.

\section{Transmission features of massive Kane fermions} 

\begin{figure*}
\centering
\begin{tabular}{cc}
(a) \qquad \qquad \qquad \qquad \qquad \qquad  \qquad \qquad \qquad& (b) \qquad \qquad \qquad \qquad \qquad \qquad \qquad \qquad \qquad \\ 
\includegraphics[trim = 0mm 0mm 0mm 0mm, scale = 0.37]{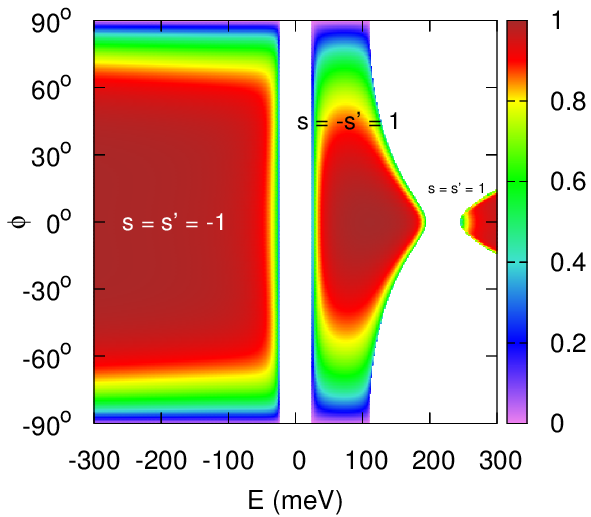} &
\includegraphics[trim = 0mm 0mm 0mm 0mm, scale = 0.37]{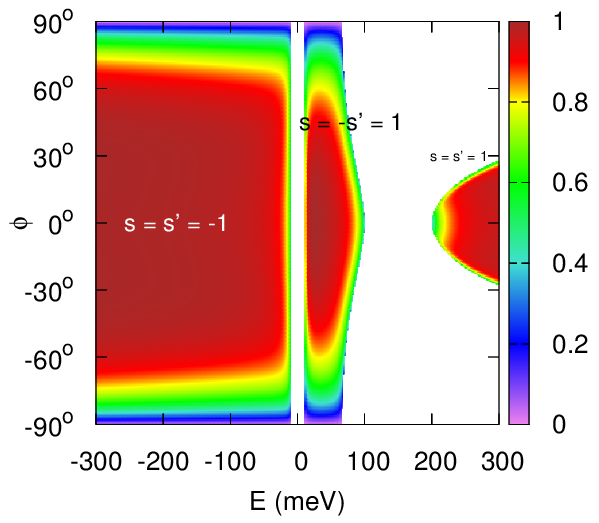}
\end{tabular}
\begin{tabular}{c}
(c) \qquad \qquad \qquad \qquad \qquad \qquad \qquad \qquad \qquad \\
\includegraphics[trim = 0mm 0mm 0mm 0mm, scale = 0.37]{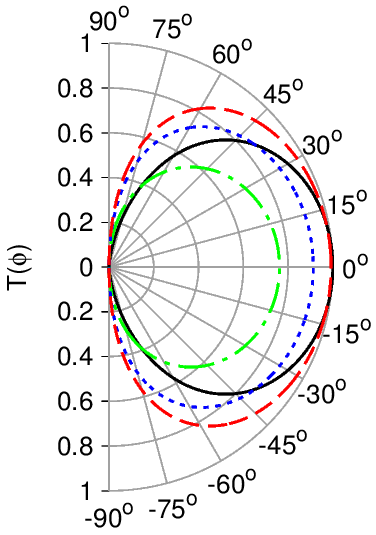}
\end{tabular}
\caption{Transmission probability as a function of the energy $E$ and the incidence angle $\phi$, for the systems (a) mass inverter using the set of values $mv^2_F = -m'v^2_F = 25$ meV and $V_0 = 220$ meV and (b) the heterojunction using $mv^2_F = 10$ meV, $m'v^2_F = -50$ meV, and $V_0 = 150$ meV. (c) Transmission probability as a function of $\phi$. KT of massive Kane fermions (black curve) is obtained in $\phi = 0^{\textrm{o}}$ for the mass inverter when $E_c = E_K = 110$ meV. Green curve illustrates how KT disappears when the value of energy is changed to $E = 30$ meV. This effect is absent for the heterojunction (blue curve) at the energy $E_c = 67$ meV given by equation \eref{FC}, where a focusing flow is obtained. Perfect transmission in this system occurs for the energy $E_K = 25$ meV (red curve) predicted by equation \eref{EK}.}
\label{flor}
\end{figure*}

In order to analyze Kane fermions transmission for MCT heterojunctions with different masses in the two regions, we plot the transmission probability $T = 1 - R$ as a function of $\phi$ and $E$ using the equations \eref{Snell} and \eref{RC} (see \fref{flor}). Two different heterojunctions are considered taking into account the restriction $V_0 \geq (|m| + |m'|)v^2_F$. The first system consists of a mass inverter which is formed by the junction between a narrow-gap semiconductor of positive mass $mv^2_F = 25$ meV and the semimetal of negative mass $m'v^2_F = -25$ meV, where the step potential is $V_0 = 220$ meV. The second system is formed by a semiconductor with $mv^2_F = 10$ meV in the region I and the semimetal with $m'v^2_F =-50$ meV in the region II, where the step potential is changed to $V_0 = 150$ meV.

\noindent In general, the heterojunctions with an abrupt step potential satisfying $|V_0| \geq (|m| + |m'|)v^2_F$ have three regimes in the transmission spectra, as shown in \fref{flor}(a) and (b). The interband transmision is located at $|m|v^2_F \leq E \leq |V_0| - |m'|v^2_F$ which is labeled using the band indices $s = -s' = 1$. The two intraband transmissions for the energy ranges $E \leq -|m|v^2_F$ and $E \geq |V_0| + |m'|v^2_F$ are denoted by $s = s' = \mp 1$, respectively. We can see that the angular wide of the transmission is decreased by the total internal reflection at the incidence range $-\phi_c \leq \phi \leq \phi_c$, where $\phi_c = \arcsin(|E - V_0|\beta'/E\beta)$ is the critical angle. This reflection appears for the intraband $s = s' = 1$ and interband transmission within the energy range $|E_c| < E < |V_0| - |m'|v^2_F$. The topology of the transmission is caused by the band gap energy regardless of the pseudo-spin structure. For instance, when the heterojunction contains a semimetal with zero mass, the intra and interband transmision regimes meet. Moreover, interband transmission can be absent if $|V_0| < (|m| + |m'|)v^2_F$ is fulfilled.  

In the two intraband transmission probabilities, as shown in \fref{flor}(a) and (b), the resemblance with massless particles is more evident when the rest mass energy becomes negligible compared with the kinetic energy. KT for normal incidence is obtained independent of the pseudo-spin structure because its conservation is nearly restored in the massless limit $|E| >> mv^2_F$. In the non-relativistic limit $|E| \approx mv^2_F$, the pseudo-spin structure has important consequences for the transmission probability. The differences begin to emerge near to the transmission gaps within the intraband transmission $s = s' = -1$. For instance, the transmission probability of massive Kane fermions is higher than in massive Dirac fermions and some pseudo-spin one configuration systems for angles of incidence far away from the normal direction \cite{Betancur,Jahani}. This is because the flat band, by located at $E = -|m|v^2_F$, improves the transmission rate. 

For interband transmission, the Kane fermion goes from the conduction band ($s = 1$) in region I to the valence band in region II ($s'= -1$), as seen in \fref{fig2}(b). We find the appearance of KT in the mass inverter (heterojunction) at the energy $E = 110$ (25) meV, which is predicted by equation \eref{EK} and shown in \fref{flor}(c). In the mass inverter, the refracted beam meets again in a symmetric spot within the region II. Whereas, for the heterojunction, the focused Kane fermion flow ocurrs at the energy $E_c = 67$ meV given by equation \eref{FC}. Both systems have a wide incidence region and energies where high transmission rate is obtained.

A comparative analysis between massive Kane fermions and pseudo-spin one particles indicates that the flat band location and pseudo-angular momentum have a correlated effect on the interband transmission. Thus, in pseudo-spin one systems can be obtained an omnidirectional perfect transmission. While in MCT crystals the massive Kane fermions perfectly tunnel for normal incidence \cite{Betancur}.

\section{Transmission through a potential barrier} 

\begin{figure}
\begin{tabular}{l}
(a)\qquad \qquad \qquad \qquad \qquad \qquad \qquad \qquad \qquad \qquad \qquad \qquad\\ 
\includegraphics[trim = 0mm 0mm 0mm 0mm, scale = 0.45]{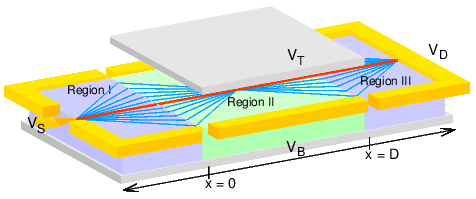} \\
(b)\qquad \qquad \qquad \qquad \qquad \qquad \qquad \qquad \qquad \qquad \qquad \qquad\\
\includegraphics[trim = 0mm 0mm 0mm 0mm, scale = 0.41]{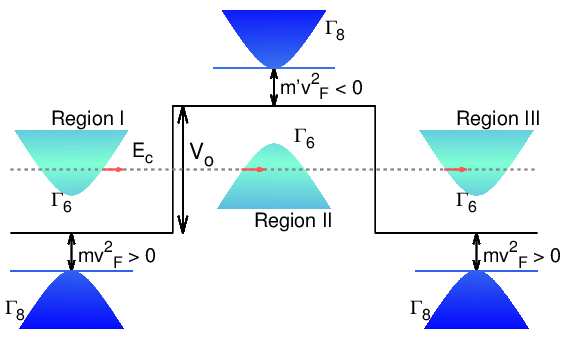}
\end{tabular}
\caption{(a) Description of the particle scattering through the mass inverter with a potential barrier created by the back-gate $V_{B}$ and top-gate $V_T$; the particle flow (blue lines) are injected from the left in the region I by the source $V_{S}$. The reflected beams are ommited and come back in the region I and II, being collected by $V_D$. The refracted beams come forward through the region II and III, where the flow is focused within and outside of the barrier. The line red indicates the appearance of Klein tunneling for normal incidence. (b) Scheme of the potential barrier and energy band structure for the heterojunction showed in (a).}
\label{Barrier}
\end{figure}

We analyze the scattering of 3D massive Kane fermions in a heterojunction formed by three MCT crystals with an abrupt potential barrier, as shown in \fref{Barrier}(a). In this system, the region I and III are formed by two identical narrow-gap semiconductors with the same negative doping level. The region II of width $D$ consists of a positively doped semimetal, where the charge carriers have opposite mass to the ones of the semiconductors for creating the mass inverter effect (see \fref{Barrier}(b)). The coherence length and mean free path are assumed to be larger than device's dimensions which are the necessary requirement of a ballistic regime. In order to describe the scattering of particles on the heterojunction, the wavefunction in region I is identical to that in equation \eref{pwr1}. The wavefunction \eref{pwr2} can be used for the region III deleting the prime on the parameters. While the wavefunction in the region II is

\begin{eqnarray}
|\Psi_{II}\rangle = & 
\frac{1}{2\sqrt{2}}a\left(\begin{array}{c}
2s'\beta'\\
\sqrt{3}\gamma'\textrm{e}^{-i\phi'}\\
-\gamma'\textrm{e}^{i\phi'}
\end{array}\right)\exp(ixp'_x/\hbar) & + \nonumber \\
&\frac{1}{2\sqrt{2}}b\left(\begin{array}{c}
-2s'\beta'\\
\sqrt{3}\gamma'\textrm{e}^{i\phi'}\\
-\gamma'\textrm{e}^{-i\phi'}
\end{array}\right)\exp(-ixp'_x/\hbar), &
\label{pwr3}
\end{eqnarray}

\noindent where $a$ and $b$ are the wavefunction's amplittudes within the barrier. Applying the boundary conditions of the Kane equation \eref{BC} at $x = 0$ and $x = D$, a coupled equation system for $r$, $a$, $b$, and $t$ is obtained
\begin{eqnarray}
s\beta(1 - r) =  s'\beta'(a - b) \nonumber\\
 w^* + wr = w'^*a + w'b \nonumber\\
s'\beta'(a\textrm{e}^{i\theta'} - b\textrm{e}^{-i\theta'}) = s\beta t\textrm{e}^{i\theta} \nonumber\\
w'^*a\textrm{e}^{i\theta'} + w'b\textrm{e}^{-i\theta'} = w^*t\textrm{e}^{i\theta},
\label{ES2}
\end{eqnarray}

\noindent where $\theta = p_xD/\hbar$ and $\theta'$ has identical form than $\theta$ in terms of $p'_x$. $w$ and $w'$ are defined as in equation \eref{ES}. Solving the $4 \times 4$ equation system \eref{ES2}, the transmission coefficient is

\begin{figure*}
\centering
\begin{tabular}{cc}
(a)\qquad \qquad \qquad \qquad \qquad \qquad \qquad \qquad \qquad& (b)\qquad \qquad \qquad \qquad \qquad \qquad \qquad \qquad \qquad\\
\includegraphics[trim = 0mm 0mm 0mm 0mm, scale = 0.35]{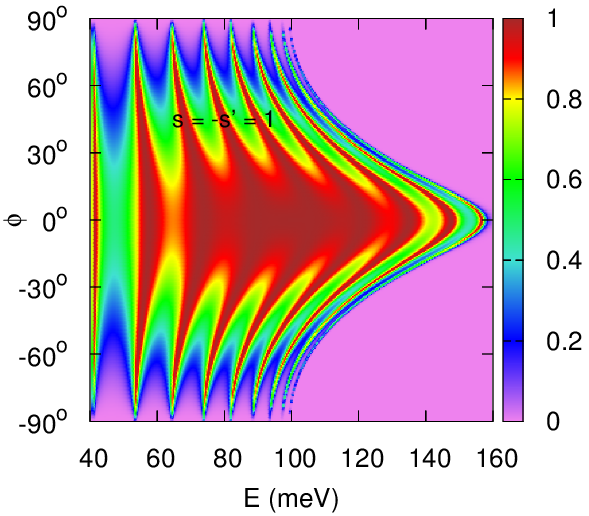}&
\includegraphics[trim = 0mm 0mm 0mm 0mm, scale = 0.35]{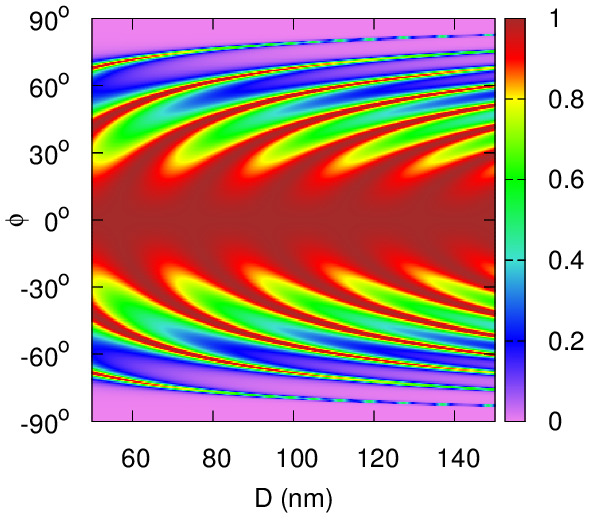}
\end{tabular}
\begin{tabular}{c}
(c)\qquad \qquad \qquad \qquad \qquad \qquad \qquad \qquad \qquad\\
\includegraphics[trim = 0mm 0mm 0mm 0mm, scale = 0.35]{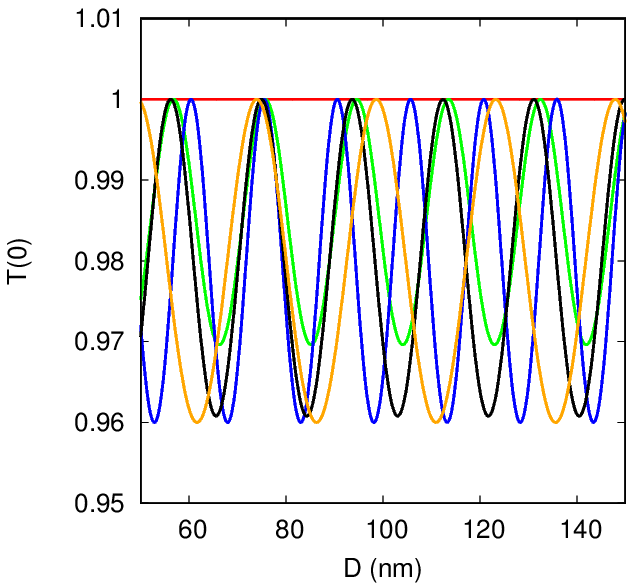}
\end{tabular}
\caption{Interband transmission probability of massive Kane fermions in a potential barrier of width $D = 100$ nm for the mass inverter, where $mv^2_F = 40$ meV ($m'v^2_F = -40$ meV) for the region I and III (II). (a) Transmission probability as a function of the energy $E$ and the incidence angle $\phi$. (b) For the value of $E = V_0/2 = 100$ meV, the transmission probability is plotted as a function of $D$ and $\phi$. (c) Klein tunneling is obtained when $E = V_0/2$ and $\phi = 0^{\textrm{o}}$ (red line). While the other curves illustrate the robustness of this effect when the incidence angle is changed to $\phi = 10^{\textrm{o}}$ (green curve). When the energies are changed to $E = 80$ meV (blue curve) and $E = 120$ meV (orange curve). Finally, if the masses are changed in the three regions of the devices with $mv^2_F = 30$ meV and $m'v^2_F = -50$ meV (black curve), where the energy $E = 96$ meV is calculated from the focusing condition \eref{FC}.}
\label{IT_barr}
\end{figure*} 

\begin{equation}
T = |t|^2 =  \frac{\cos^2\phi\cos^2\phi'}{\cos^2\phi\cos^2\phi'\cos^2\theta' + f^2(\phi,\phi')\sin^2\theta'},
\label{TKF}
\end{equation}

\noindent where the function $f(\phi,\phi')$ is given by

\begin{eqnarray}
f(\phi,\phi') & = &\frac{1}{2}\bigg[\frac{\gamma\beta'}{\gamma'\beta}\left(\cos^2\phi + \frac{1}{4}\sin^2\phi\right) + \nonumber \\
& &\frac{\gamma'\beta}{\gamma\beta'}\left(\cos^2\phi' + \frac{1}{4}\sin^2\phi'\right) - \frac{1}{2}ss'\sin\phi\sin\phi'\bigg]. \nonumber \\
& & 
\label{la_f} 
\end{eqnarray}

In order to plot the transmission probability for interband transmission shown in \fref{IT_barr} (a), we set the parameters $mv^2_F = 40$ meV, $m'v^2_F = -40$ meV, $D = 100$ nm, and $V_0 = 200$ meV as input for the expressions \eref{TKF}, \eref{la_f}, and \eref{Snell}. For any incidence angle, the Fabry-P\'erot fringes emerge as a function of $E$, $V_0$ and $D$. The perfect transmission by constructive interference of massive Kane fermions within barrier are obtained when the condition $p'_xD = n\pi\hbar$ is satisfied. This resonance condition has remained unchanged for all the studied scattering problems of relativistic particles in a potential barrier, since it essentially depends on the wave behavior of the particle. Let us note that non-resonant perfect transmission seems indistinguishable from resonant one in \fref{IT_barr} (a). However, for the value $E = V_0/2$ and observing how the transmission probability changes as a function of $D$ and $\phi$ in \fref{IT_barr} (b), we can see that for normal incidence non-resonant perfect transmission is obtained. This can be explained from the transmission coefficient of massive Kane fermions in equation \eref{TKF}, which looks very similar to that of massless Dirac fermions, with a different form of $f(\phi,\phi')$ \cite{Katsnelson}. The non-resonant perfect transmission under normal incidence depends exclusively on the value of $f(0,\pi)$. Thus, electrons in graphene impinging on the barrier perfectly cross without resonance for any value of $E$ and $V(x)$ because $f(0,\pi) = f(0,0) = 1$, leading to the well-known KT of massless Dirac fermions \cite{Katsnelson}. For the special value of $E = V_0/2$ in equation \eref{la_f} gives $f(0,\pi) = 1$ confirming that the perfect transmission of massive Kane fermions does not correspond to constructive interference. This result differs of massive Dirac fermions, where perfect non-resonant tunneling never occurs for interband transmission \cite{Jahani}. Therefore, the perfect transmission of massive Kane fermions is not destroyed considering the scattering in a potential barrier because the mass inverter effect restores the conservation of $J_x$.

The robustness of the perfect transmission of massive Kane fermions is examined in situations where the experimental setup deviates from ideal conditions. If the energy changes around 20 \% regarding the value of focusing \eref{FC}, resonant transmission peaks appear with a tiny amplitude and the deviation of perfect transmission does not exceed the 4 \%, as shown in \fref{IT_barr}(c). Similar deviation of transmission and oscillations are expected when the masses among regions are changed by $mv^2_F = 30$ meV and $mv^2_F = -50$ meV or if the particles impinging on the barrier have an incidence angle of $\phi = \pm 10^{\textrm{o}}$. These observations indicate that the perfect transmission of 3D massive Kane fermions in mass inverters of MCT crystals is quite robust and can be observed over a wide operation range. We note that if the masses are changed, the KT must be observed for the value of energy $E_K$ predicted by equation \eref{EK}.

\section{Conclusions and final remarks} \label{cr}

In summary, we have discussed the transmission properties of massive Kane fermions in MCT heterojunctions. We found that the perfect transmission of massive Kane fermions can be obtained when the heterojunction is designed with a narrow-gap semiconductor and semimetal forming the mass inverter. This effect, which is due to the heterojunction of relativistic materials with inverted band order, is identified as the main factor for a perfect transmission of massive particles. For general heterojunctions, perfect transmission can emerge for a special value of the doping level. We identified the common features of the transmission properties of massive Kane fermions with any pseudo-relativistic particle, where KT is always restored in the massless limit regardless of the pseudospin structure. Under the focusing condition $E = V_0/2$, the mass inverter operates as a Veselago lens allowing the conservation of $J_x$. Perfect transmission for normal incidence is not destroyed by considering scattering of massive Kane fermions in a potential barrier. Such phenomena is hardly affected by possible experimental deviations. Our findings may have importance for the development of nano-electronic devices due to the tunable band gap of MCT further that several transmission features of massless fermions are recovered. MCT heterojunctions can be used to make Veselago lenses. For instance, mass inverter could be used as probing tip of an enhanced scanning tunneling microscope. On the other hand, the large number of experimental studies on MCT crystals leads us to expect that our results may have concrete applications in electron optics for the nearest future.

\ack{
One of the authors (Y.B-O.) gratefully acknowledge a SNI schoolarship from Consejo Nacional de Ciencia y Tecnolog\'ia of M\'exico (Conacyt-M\'exico). V. Gupta would like to thank Conacyt-M\'exico for its support.}

\appendix
\setcounter{section}{1}
\section*{Appendix A: abrupt step potential condition and characteristic lengths}
We establish the condition of abrupt step potential for massive pseudo-relativistic particles using a semiclassical approximation. Smooth step potential can be expressed as $V(x) = Fx - V_0/2$, where $F = V_0/2d$ for $|x| \leq d$. While it is zero ($V_0$) when $x < -d$ ($x > d$). Considering particles impinging the potential with energy $E_F = V_0/2$, the evanescent modes occurs for imaginary values of $k'_x(x)$. Using the conservation of linear momentum $\hbar k_y$, $\hbar k_z$, and energy we obtained 

\begin{equation}
k'^2_x(x) = \overline{F}^2x^2 - \overline{k}^2 < 0,
\end{equation}

\noindent where $\overline{F} = F/\hbar v_F$ and $\overline{k}^2 = k^2_y + k^2_z + k^2_B$ being $k_B = mv_F/\hbar$ the de Broglie wave vector. Hence, the transmission by evanescent modes appear in the range $|x| < l = \overline{k}/\overline{F}$. In the semiclassical approximation the transmission probability is given by $T \approx \textrm{e}^{-2S}$, where the action $S$ is written as

\begin{equation}
S = \int^l_{-l} |k'_x|dx = \overline{k}l\int^1_{-1} \sqrt{1 - u^2}du = \pi\overline{k}^2/2
\overline{F}.
\end{equation}

\noindent In terms of the Fermi wave vector $k_F = \beta E_F/\hbar v_F$ with $\beta = [1 - (mv^2_F/E_F)^2]^{1/2}$, the following quantities can be expressed as $k^2_y + k^2_z = k^2_F\sin^2\phi$ and $\overline{F} = k_F/\beta d$. Thus, the transmission probability is

\begin{equation}
T(\phi) \approx \exp\left(-\frac{\pi\beta k^2_Bd}{k_F}\right)\exp(-\pi\beta k_Fd\sin^2\phi).
\end{equation}  

\noindent We note that the transmission probability for the case of massless particles is recovered \cite{Low2,Fuchs,Ghosh}. High transmission via evanescent modes is guaranteed when the condition of abrupt step potential $\beta k^2_Bd/k_F < 1$ and $\beta k_Fd < 1$ is fulfilled. An equivalent sharpness criterion can be expressed as $d < d_c = \textrm{min}\{d_1,d_2\}$, where $d_1 = 2\hbar v_F/\beta^2V_0$ and $d_2 = \hbar V_0/2m^2v^3_F$. With the set of values $mv^2_F = 40$ meV, $V_0 = 100$ meV and $\hbar v_F = 704.3$ meV$\cdot$nm, the step potential can be considered as abrupt if $d < 22$ nm. The mean free path of electrons in MCT crystals is given by $\lambda = \mu \hbar k_F/e$, where $e$ is the charge of electron and the charge mobility $\mu$ = 10$^5$ cm$^2$/(V$\cdot$s) which is experimentally estimated at $T = 1.4$ K \cite{Noel}. For the same device, we found that $\lambda = 0.3 \ \mu$m. 

\section*{Appendix B: conservation of $J_x$ in MCT heterojunctions}

We show the appearence of KT of massive Kane fermions by the conservation of $J_x$. This issue is discussed following a similar line of thinking as for massless Dirac fermions \cite{Fuchs}. The Heisenberg equation of motion allows to obtain the velocity operator as

\begin{equation}
\vec{v} = -\frac{i}{\hbar}[\vec{r},H] = v_F\vec{J}.
\end{equation}

\noindent Since $p_y$ is a conserved quantity, a state $|\psi(t)\rangle$ with normal incidence $(\phi = 0)$ has zero momentum along $y$ at any time $t$. Then, the expectation value of $v_x$ using the wavefunction \eref{WM1} is   

\begin{equation}
\langle \psi_s|v_x|\psi_s\rangle = v_F\langle \psi_s|J_x|\psi_s\rangle = v_F\beta\gamma.
\end{equation}

\noindent When the normal incident massive Kane fermion is scattered by the barrier or potential step in a time $t_s$, the expectation value of $v'_x$ for the transmitted state in the region II has a similar relation

\begin{equation}
\langle \psi_{s'}|v'_x|\psi_{s'}\rangle = v_F\beta'\gamma'.
\end{equation}

\noindent These group velocities $\langle v_x \rangle$ and $\langle v'_x \rangle$ are the same under the condition $\beta\gamma = \beta'\gamma'$, which is always satisfied for the mass inverter when the particle flow is focused. In a general heterojunction, incident and transmitted states have the same group velocity for the special energy value given by equation \eref{EK}.

\end{document}